%% file: main.tex
\documentclass[sigconf, nonacm]{acmart}

\usepackage{graphicx}
\usepackage{balance}  

\usepackage{enumitem}
\usepackage{tikz}
\usepackage{xcolor}
\newcommand*\circled[1]{\tikz[baseline=(char.base)]{
           \node[shape=circle,fill,inner sep=1pt] (char) {\textcolor{white}{#1}};}}
\usepackage{color}
\usepackage{multirow}
\usepackage{subfigure}
\usepackage[export]{adjustbox}
\usepackage{enumitem}
\usepackage{subcaption}
\usepackage{hhline}
\usepackage{balance}
\usepackage{colortbl}
\usepackage{units}
\usepackage{marginnote}
\usepackage{pgfplotstable,array}
\usepackage{marginnote}
\usepackage{float}
\usepackage{pgfplots}
\usepackage{pgf,tikz}
\usepackage{mathtools}
\usepackage{xstring}

\usepackage{needspace}
\usepackage{balance}

\usepackage{enumitem}
\setlist{noitemsep, topsep=0pt, parsep=0pt, partopsep=0pt}

\usetikzlibrary{spy}
\usetikzlibrary{patterns}

\newcommand{\stitle}[1]{\vspace{0.8ex}\noindent\textup{\textbf{#1}}}

\usepackage{algorithmicx}
\usepackage{algorithm}
\usepackage{algpseudocode}
\usepackage{listings}

\newcommand{\rev}[1]{#1}

\newcommand{\hide}[1]{}

\AtBeginDocument{%
  }

\lstdefinestyle{sql}{
    language=SQL,
    basicstyle=\ttfamily\small,
    keywordstyle=\color{blue}\bfseries,
    stringstyle=\color{red},
    commentstyle=\color{gray}\itshape,
    numbers=left,
    numberstyle=\tiny\color{gray},
    stepnumber=1,
    numbersep=10pt,
    backgroundcolor=\color{white},
    showspaces=false,
    showstringspaces=false,
    showtabs=false,
    frame=single,
    rulecolor=\color{black},
    tabsize=2,
    captionpos=b,
    breaklines=true,
    breakatwhitespace=false,
    escapeinside={(*@}{@*)},
    morekeywords={SELECT, FROM, WHERE, ORDER, BY, AS, INSERT, INTO, VALUES, CREATE, TABLE, PRIMARY, KEY, FOREIGN, REFERENCES, INDEX, AUTO_INCREMENT, VARCHAR, INT, NOT, NULL, MAX, SUM, GROUP, AND, OR, INNER, JOIN, ON}
}

\makeatletter
\newcommand{\currentfsize}{\f@size pt}
\makeatother

\newdimen\fsize

\newcommand\vldbdoi{XX.XX/XXX.XX}
\newcommand\vldbpages{XXX-XXX}
\newcommand\vldbvolume{14}
\newcommand\vldbissue{1}
\newcommand\vldbyear{2020}
\newcommand\vldbauthors{\authors}
\newcommand\vldbtitle{\shorttitle} 
\newcommand\vldbavailabilityurl{URL_TO_YOUR_ARTIFACTS}
\newcommand\vldbpagestyle{plain} 

\begin{document}
\title{\rev{Toward Temporal Attribution Analytics in Dataflows}} \subtitle{{\bf [Vision
  Paper]}}

\author{Chrysanthi Kosyfaki}
\affiliation{%
  \institution{Hong Kong University of Science and Technology}
  \city{Hong Kong SAR}
  \state{China}
}
\email{ckosyfaki@cse.ust.hk}

\author{Ruiyuan Zhang}
\affiliation{%
  \institution{Hong Kong Generative AI R\&D Center}
  \city{Hong Kong SAR}
  \state{China}
}
\email{zry@hkgai.org}

\author{Nikos Mamoulis}
\affiliation{%
  \institution{University of Ioannina}
  \city{Ioannina}
  \state{Greece}
}
\email{nikos@cs.uoi.gr} 

\author{Xiaofang Zhou}
\affiliation{%
  \institution{Hong Kong University of Science and Technology}
  \city{Hong Kong SAR}
  \state{China}
}
\email{zxf@cse.ust.hk}

\begin{abstract}
Data provenance (the process of determining the origin and derivation of 
data outputs) has applications across multiple domains including explaining 
database query results and auditing scientific workflows. Despite decades 
of research,
provenance tracing
remains challenging due to its high computational cost and storage requirements. In 
streaming systems such as Apache Flink, \rev{fine-grained} provenance graphs can grow 
super-linearly with data volume, posing significant scalability challenges.
\rev{We define temporal attribution, a new lightweight form of provenance, appropriate for certain tasks, such as  monitoring dependencies between system components over time quantitatively.} 
\rev{Temporal attribution enables time-focused analysis that does not require fine-grained, tuple-level dependency meta-data.} 
\rev{Inspired by volume-based provenance tracking in Temporal Interaction Networks 
(TINs)}, we 
demonstrate TINs' 
applicability in \rev{succinctly modeling  quantified
data exchanges between dataflow operators} in stream data processing systems \rev{and in processing workflows, in general, over time}. 
We classify data into discrete and liquid types, define 
five temporal provenance query types, and propose a state-based indexing 
approach. Our vision outlines research directions toward making \rev{this new form of temporal attribution} a practical tool for large-scale dataflow \rev{analytics}.
\end{abstract}

\maketitle

\pagestyle{\vldbpagestyle}
\begingroup\small\noindent\raggedright\textbf{PVLDB Reference Format:}\\
\vldbauthors. \vldbtitle. PVLDB, \vldbvolume(\vldbissue): \vldbpages, \vldbyear.\\
\href{https://doi.org/\vldbdoi}{doi:\vldbdoi}
\endgroup
\begingroup
\renewcommand\thefootnote{}\footnote{\noindent
This work is licensed under the Creative Commons BY-NC-ND 4.0 International License. Visit \url{https://creativecommons.org/licenses/by-nc-nd/4.0/} to view a copy of this license. For any use beyond those covered by this license, obtain permission by emailing \href{mailto:info@vldb.org}{info@vldb.org}. Copyright is held by the owner/author(s). Publication rights licensed to the VLDB Endowment. \\
\raggedright Proceedings of the VLDB Endowment, Vol. \vldbvolume, No. \vldbissue\ %
ISSN 2150-8097. \\
\href{https://doi.org/\vldbdoi}{doi:\vldbdoi} \\
}\addtocounter{footnote}{-1}\endgroup


\input{introduction}

\input{overview}

\input{tins}
\input{solutions}

\input{temporalindex}
\input{relatedwork}
\input{next}
\input{conclusions}

\bibliographystyle{ACM-Reference-Format}
\bibliography{references}

\end{document}

%% file: introduction.tex
\section{Introduction}\label{sec:intro}
Data provenance (also known as data lineage) refers to the process of 
identifying the origin and transformations of data throughout its lifecycle
\cite{DBLP:conf/icdt/BunemanKT01, DBLP:conf/pods/BunemanKT02,
  almuntashiri2025using, almuntashiri2024llms}.
It is a
fundamental
concept in modern 
data management, enabling transparency, trust, and accountability in 
data-driven systems \cite{lee2013spatio, green2010provenance}.
In relational databases
\cite{buneman2007provenance, tan2007provenance,
  senellart2019provenance, green2017semiring, sen2025provsql,
  kementsietsidis2009provenance}, 
provenance
can
explain the results of complex SQL queries, 
supportiing query debugging \cite{niu2017debugging,
  whittaker2018debugging, de2014debugging, zhao2020debugging, zipperle2022provenance, lucia2015data,
  wu2014diagnosing, DBLP:conf/icde/MullerE22}, view maintenance, and fine-grained access control. In 
distributed and streaming systems \cite{DBLP:conf/debs/GlavicEFT13,palyvos2018genealog,DBLP:journals/pvldb/Palyvos-Giannas20,DBLP:journals/pvldb/Palyvos-Giannas22, chen2017distributed}, it
helps model and trace
large-scale streaming dataflows for 
fault recovery and performance optimization. Its importance extends to 
multiple domains: in financial networks, provenance helps detect illicit 
activities and trace suspicious transactions; in cybersecurity, it identifies 
malicious behaviors linked to IP addresses; in healthcare, it ensures 
compliance and reproducibility by tracking the sources of clinical
data; in AI, it can validate outputs generated by
large language models (LLMs) and audit 
workflows in scientific experiments.

\rev{Fine-grained provenance, the most detailed form of data provenance, identifies which source tuples contributed to a given output of a workflow (and how) \cite{DBLP:journals/pvldb/Ruan0DLOZ19, chapman2024supporting, DBLP:conf/debs/GlavicEFT13,DBLP:conf/icde/MullerE22,DBLP:conf/btw/GlavicEFT11, DBLP:journals/pvldb/ZhengI20,DBLP:conf/edbt/MissierPB10,DBLP:conf/sigmod/Yao22,DBLP:conf/dmsn/HuqWA10, palyvos2018genealog, DBLP:journals/pvldb/Palyvos-Giannas22, DBLP:journals/pvldb/Palyvos-Giannas20, DBLP:journals/pacmmod/YamadaKSAM25, DBLP:conf/vldb/GreenKIT07,DBLP:conf/icde/NiuKGGLR17,DBLP:journals/pvldb/PsallidasW18,DBLP:conf/cidr/Widom05,DBLP:conf/icde/Woodruff97}.}
Despite decades of research, efficiently tracking and storing \rev{fine-grained} provenance 
information remains a major challenge, due to its
high computational 
cost and storage overhead.
For instance, consider 
a provenance graph that captures record-level lineage in a modern distributed 
streaming system such as Apache Flink or Spark. Unlike the static job graph, 
the provenance graph expands as data flows through the system. Its size can 
grow super-linearly
with data volume 
especially for operations like joins and aggregations \cite{interlandi2015titian}. 
This growth introduces
severe
scalability challenges in memory/storage and network traffic;
the latency of lineage 
tracking can be too high in environments 
processing millions of events per second.
\rev{Overall, fine-grained provenance is space demanding, expensive to track and could be overwhelming to users, especially in workflows that process huge data volumes and need to be monitored for long time periods.
In addition, tuple-level provenance may also be the wrong unit of information when comparing provenance between outputs that are produced at different times; the tuples that contributed to these outputs could be totally different, so the only way to compare them is based on the sources (or paths) and the volume of data from them that affected the outputs. 
At the other extreme, coarse-grained lineage systems (i.e., table-level or partition-level lineage used for data governance) operate at far too coarse a granularity \cite{psallidas2023oneprovenance, jacques2025unified}. They track which datasets feed which datasets across an organization, 
but they do not quantize the contribution of a dataset to a given output nor do they consider the data processing workflow.}

\rev{Motivated by the above, we propose the new concept of {\em temporal attribution}, which computes \textit{how much} each of the dataflow sources contributed to a specific output. In its generalized form, attribution quantizes the contribution of sources to \textit{groups of output tuples}. By comparing this contribution over time, we can identify changes in system behavior and monitor load distribution between operators and processing paths.
Attribution has already been considered as a tool  for interpreting neural network results 
\cite{ZhengZXCC24,walker2025explainingreasoninglargelanguage}; we are the first to suggest its use for the outputs of 
streaming dataflows.}
\rev{Previous work on {\em temporal provenance}} \cite{wu2019zeno, beheshti2012temporal, chen2012temporal, reha2023anomaly,
zhou2012distributed} 
attaches timestamps to
either \rev{coarse-grained} or \rev{fine-grained} provenance information, capturing when derivations 
occurred. This allows time-focused provenance representation
and tracking, 
\rev{e.g.,} ``which sources 
contributed to this output between 2PM and 3PM?'', \rev{however, it does not address temporal attribution (e.g., how much each source contributed)}.

\rev{To facilitate temporal attribution, inspired by previous work on network flow analytics in Temporal Interaction 
Networks (TINs)} \cite{DBLP:conf/edbt/KosyfakiMPT19,DBLP:conf/icde/KosyfakiMPT21,DBLP:conf/icde/KosyfakiM22}, \rev{we propose modeling
workflows as TINs, where nodes exchange (and possibly transform) data over time. Since our focus is quantitative provenance (i.e., attribution), we discretize time, and consider the workflow operators as processing groups of tuples in regular time intervals (e.g., processing is summarized for every period of 1 second), instead of monitoring the processing of individual tuples at a fine time granularity.}
\rev{This abstraction of the dataflow as a TIN facilitates the temporal attribution tracking that we are targeting and reduces the necessary meta-data required for it.
Modeling the dataflow as a TIN achieves state-based compression at its core:
instead of recording per-tuple dependencies, we 
we do this between temporal states
that aggregate data over time windows.
This exploits the observation that, in many streaming
workloads, large data volumes are transferred without materially
changing the aggregate attribution structure (e.g., during window
accumulation). By representing such periods with a small number
of states, we achieve substantial compression while still preserving sufficient information to answer temporal attribution queries,
without reconstructing fine-grained histories.
Attribution is especially appropriate in cases where data are {\em liquid}, i.e., become indistinguishable when merged or split.
Such data are natural in some applications (e.g., financial workflows where nodes exchange liquid assets), while data anonymization may also be enforced in distributed environments where data privacy is a concern \cite{zhang2017privacy,politou2022privacy,pan2023data,zhang2023differentially}}. 

\noindent\textbf{Our contributions include:} 
\circled{1} demonstrating how TINs provide a model for \rev{dataflows that facilitates temporal attribution} across applications \rev{that process data in a workflow};
\circled{2} classifying data into discrete\rev{/identifiable} types (where transferred data maintain 
identity) versus liquid types (where quantities \rev{become indistinguishable after merging});
\circled{3} formalizing five temporal attribution query types: backward \rev{attribution} 
(where-from), forward \rev{attribution} (where-to), temporal lineage 
(when-contributed), flow lineage (how-much-through), and versioning 
\rev{attribution} (how-changed) that leverage TINs' temporal structure; and 
\circled{4} proposing temporal \rev{attribution} indexing based on vertex state sequences 
that enables efficient query evaluation without reconstructing entire 
interaction histories. By doing so, this agenda aims to help researchers 
develop scalable solutions for efficiently tracking \rev{temporal attribution}.

\noindent\textbf{Roadmap:} 
The rest of the paper is organized as follows:
Section \ref{sec:concepts} provides an overview of data
provenance and TINs. Section \ref{sec:tin} discusses TINs and demostrates their
application to modeling \rev{dataflows}. Section \ref{sec:data}
analyzes different data classes and their impact on \rev{temporal attribution} tracking. Section \ref{sec:tpi} presents our temporal \rev{attribution indexing and queries}. Section \ref{sec:relwork} reviews related work. In Section
\ref{sec:next} we propose a research agenda and Section
\ref{sec:conclusions} concludes the paper.

%% file: overview.tex
\vspace{-5pt}
\section{Data Provenance and TINs}\label{sec:concepts}
This section provides the necessary background. We 
begin with the classic provenance models and their limitations in 
dynamic settings, then present the Temporal Interaction Network 
formalism from prior work that we build upon.

\subsection{Data Provenance}
Data provenance captures the origin and derivation of data. The 
seminal work by Buneman et al. \cite{DBLP:conf/icdt/BunemanKT01,
  DBLP:conf/pods/BunemanKT02, green2017semiring} defined three core provenance semantics for 
relational databases: \textit{Where-provenance} identifies which input tuples 
contributed to an output tuple. For a query result, 
where-provenance returns the set of source tuples that appear in at 
least one derivation of the result. \textit{Why-provenance} identifies which input tuples are 
necessary for an output. It returns the minimal sets of source 
tuples sufficient to derive the output, corresponding to witness 
bases. \textit{How-provenance} provides an algebraic expression 
showing how output values depend on input values. Green et al.
\cite{green2007provenance} formalized the concept of {\em provenance 
semirings}, which annotate tuples with polynomial expressions tracking 
their derivation.

These models were designed for \textit{snapshot queries} over static 
databases, where provenance can be computed by inspecting the query 
plan and tracing tuple dependencies. However, they face fundamental 
challenges in \textit{continuous} systems.
\rev{First,}
traditional provenance does not 
capture \textit{when} data contributions occurred; e.g., 
in a streaming 
system,
where-provenance cannot distinguish between contributions at different
time periods.
\rev{Second,} 
they do not handle operators that maintain 
state across multiple inputs, such as windowed aggregations or 
stateful joins common in stream processing.
\rev{Third, quantities are not tracked.}
In 
systems where data 
represents quantities (event counts, monetary values, network 
volumes), provenance must track \textit{how much} each source 
contributed, not just \textit{whether} it contributed.
\rev{The above limitations can be addressed by fine-grained provenance mechanisms for stream data processing systems, like Ariadne \cite{DBLP:conf/debs/GlavicEFT13} and 
Genealog \cite{palyvos2018genealog}; however, at the cost of tracking tuple-level dependencies between produced outputs throughout the workflow.
}


\subsection{Temporal Interaction Networks}
Temporal Interaction Networks (TINs) are a graph-based formalism 
for modeling time-varying data flows \cite{DBLP:conf/icde/KosyfakiM22,
  DBLP:conf/icde/KosyfakiMPT21}. A TIN is a triple $G = (V, E, R)$ where $V$ is a 
set of vertices, $E \subseteq V \times V$ is a set of directed 
edges, and $R$ is a set of interactions. Each interaction 
$r \in R$ is a quadruple $(r_s, r_d, r_t, r_q)$ with source vertex 
$r_s$, destination vertex $r_d$,
\rev{such that $(r_s,r_d)\in E$}, timestamp $r_t$, 
and transferred quantity $r_q \in \mathbb{R}^+$. Each vertex 
maintains a time-varying buffer $B_v(t)$ representing the accumulated 
quantity in $v$ at time $t$.
Interactions increase destination buffers and decrease 
source buffers, enabling TINs to model both transient flows and 
accumulation.

TINs differ from traditional temporal graphs in three key aspects: 
\circled{1} \textit{explicit quantity tracking}: each interaction 
transfers a specific quantity, it is not just a binary 
connection or an event occurrence; \circled{2} \textit{buffer state}: 
vertices maintain accumulated quantities over time, capturing 
stateful behavior; \circled{3} \textit{flow semantics}: interactions 
represent data transfers, where a quantity 
leaves one vertex and arrives at another.

These properties make TINs well-suited for \rev{modeling dataflows for temporal attribution tracking}.
The 
temporal dimension captures \textit{when} data flows occurred, the 
\rev{quantity} dimension captures \textit{how much} data flowed, and 
the buffer mechanism captures stateful operators common in streaming 
systems. 
\rev{Ref. \cite{DBLP:conf/icde/KosyfakiM22} shows how to propagate meta-data in TINs for provenance tracking of {\em liquid} data (e.g., money) in financial exchange networks, under certain models governing data selection from buffers for propagation.} 
In the following sections, we demonstrate how to leverage 
TINs for efficient \rev{temporal attribution representation and tracking in data stream systems.}

%% file: tins.tex
\needspace{4\baselineskip}
\section{TINs for Temporal Attribution}\label{sec:tin}
\rev{TINs can be used to model dataflows in a way that facilitates temporal attribution. Specifically, 
we model each dataflow operator as a node in a TIN and 
summarize processing in regular time intervals (e.g., every second).
We consider aggregate data transfers between connected operators in consecutive time windows and each such aggregate transfer is represented as a TIN interaction ($src, dst, t, q$), where $q$ is the number of tuples (i.e., aggregated quantity) exchanged. Operators are classified based on how they transform incoming interactions into outgoing ones: (i) stateless operators (e.g., map/filter) apply a transformation function that preserves or scales quantities (e.g., filtering reduces $q$); (ii)  partitioning operators (e.g., keyBy/shuffle) split incoming quantities (groups of tuples) to  multiple outgoing interactions whose quantities
sum to the input; and (iii) stateful operators (e.g., windows/joins) accumulate incoming quantities within time windows and emit new quantities when each window triggers, reflecting aggregated outputs. Attribution meta-data follow these transformations by propagating and aggregating quantities across interactions and by tracking dependencies between them, thus, forming an {\em attribution graph}.} 

\begin{figure}[!ht] 
\centering
\setlength{\tabcolsep}{2pt} 

\begin{tabular}{cc}
\includegraphics[width=0.40\columnwidth]{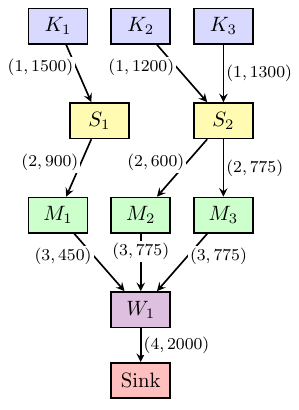} &
\includegraphics[width=0.62\columnwidth]{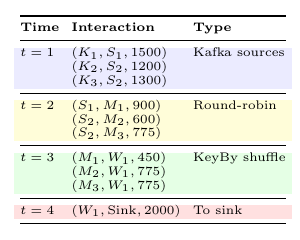} \\[3pt] 
(a) TIN & (b) Interactions \\
\end{tabular}

\vspace{10pt} 
\caption{TIN-based provenance framework.}
\label{fig:overview}

\vspace{15pt} %
\end{figure}

\rev{As an example of how to model a dataflow as a TIN, consider a Flink job, illustrated in Figure \ref{fig:overview}(a) that processes user activity and transactions streams from an e-commerce platform to compute the total number of items purchased per product over tumbling windows. Nodes inlcude Kafka 
sources ($K_1, K_2, K_3$), source operators ($S_1, S_2$), map operators 
($M_1, M_2, M_3$), a window operator ($W_1$), and a sink.}
\rev{The system ingests three event streams from Kafka partitions $K_1$, $K_2$, $K_3$: ViewEvent: (userID, productID, timestamp) capturing product page visits,  AddToCartEvent: (userID, productID, quantity, timestamp) capturing cart additions, and PurchaseEvent: (userID, productID, quantity, price, timestamp) capturing completed purchases. Source operators $S_1$, $S_2$ ingest and unify these streams, after which map operators transform events into quantitative records: $M_1$ filters view events and emits non-contributing signals, while $M_2$ and $M_3$ map add-to-cart and purchase events, respectively, into (productID, quantity) pairs. These records are then partitioned by key and routed to a window operator $W_1$, which aggregates the total quantity per product within each time window and emits the result to the sink.}

Figure \ref{fig:overview}(b) shows TIN interactions ordered by \rev{discretized/abstracted} time. At \rev{time window} $t=1$, 
Kafka sources ingest events: $(K_1, S_1, 1500)$, $(K_2, S_2, 1200)$, 
$(K_3, S_2, 1300)$.
At $t=2$, source operators apply round-robin 
partitioning: $(S_1, M_1, 900)$, $(S_2, M_2, 600)$, $(S_2, M_3, 775)$. 
At $t=3$, map operators perform KeyBy shuffle to the window: 
$(M_1, W_1, 450)$, $(M_2, W_1, 775)$, $(M_3, W_1, 775)$. At $t=4$, the 
window fires and sends aggregated results to the sink: 
\linebreak $(W_1, \text{Sink}, 2000)$.

During the time interval $[3,4)$, $W_1$ receives 2000 \rev{tuples} from 
$M_1, M_2, M_3$. Instead of storing \rev{individual provenance back-links} for all \rev{these tuples} 
we \rev{store three attribution back-links to the interactions of the three mappers: $(M_1,3,450)$, $(M_2,3,775)$, and $(M_3,3,775)$}.
This {\em state-based compression} is particularly effective 
for windowed aggregations where thousands of events arrive between window 
boundaries but produce only a few state transitions.
\rev{For operator-internal logic (e.g., auditing which specific tuples were removed by filtering or identifying the exact set of records merged during an aggregation), traditional fine-grained provenance remains necessary. Our model 
does not track the identity of individual records within the dataflow, but
captures (i) the resulting changes in data volume and flow rates and (ii) aggregate dependencies between states of (consecutive) operators.
This is done by linking consecutive interactions at different nodes if they correspond to the same groups of tuples. For example, 
$(K_1, S_1, 1500)$ is linked to $(S_1, M_1, 900)$. }

\noindent \textbf{TINs vs. Traditional Provenance Graphs.}
\rev{
When the goal is temporal attribution tracking, TINs present key advantages in modeling dataflow graphs (at a coarse time granularity).}
\rev{First, although quantitative information 
can be aggregated from   
contribution graphs used by fine-grained provenance systems such as 
Genealog \cite{palyvos2018genealog}, this bears substantial computational cost. Second, due to the super-linear space complexity of contribution graphs, we cannot afford to keep them for a long time, for the sake of temporal quantitative dependency analytics over historical data.} 
\rev{Even with state-of-the-art optimizations designed to minimize per-tuple metadata \cite{DBLP:journals/pvldb/Palyvos-Giannas20}, the number of nodes becomes too high to store, making long-term monitoring unfeasible. Lazy fine-grained provenance schemes \cite{DBLP:conf/debs/GlavicEFT13,DBLP:journals/pacmmod/YamadaKSAM25} have reduced cost, but also require partially replaying the execution of the stream.}
TINs 
encode time in each \rev{coarse-level} interaction, with our vision focusing on enabling direct index-based
retrieval. 
Second, \rev{traditional fine-grained provenance graphs do not scale  with data volume; for instance, a Flink job processing $1$ million events per second would generate $3.6$ billion provenance nodes per hour at the ingestion layer alone. Even with state-of-the-art optimizations designed to minimize per-tuple metadata \cite{DBLP:journals/pvldb/Palyvos-Giannas20}, the number of nodes becomes too high to store, making long-term monitoring unfeasible.}
Second, \rev{traditional fine-grained provenance graphs do not scale  with data volume; for instance, a Flink job processing $1$ million events per second would generate $3.6$ billion provenance nodes per hour at the ingestion layer alone. Even with state-of-the-art optimizations designed to minimize per-tuple metadata \cite{DBLP:journals/pvldb/Palyvos-Giannas20}, the number of nodes becomes too high to store, making long-term monitoring unfeasible.}
Second, \rev{traditional fine-grained provenance graphs do not scale  with data volume; for instance, a Flink job processing $1$ million events per second would generate $3.6$ billion provenance nodes per hour at the ingestion layer alone. Even with state-of-the-art optimizations designed to minimize per-tuple metadata \cite{DBLP:journals/pvldb/Palyvos-Giannas20}, the number of nodes becomes too high to store, making long-term monitoring unfeasible.}
TINs compress flows 
into temporal states: instead of 
numerous of individual \rev{tuples}, we store 
aggregated interactions. 
Third, TINs explicitly capture quantities, enabling \rev{attribution-based analysis}. In the Flink example, we can \rev{use the group-to-group quantity links to answer}
 ``How much data 
flowed from $K_2$ to $W_1$?''. 
\rev{When detailed contribution graphs \cite{palyvos2018genealog} are used},
this requires \rev{traversing numerous paths.}

\noindent \textbf{Limitations of State Compression.}
While state-based compression may achieve dramatic \rev{space reduction}
for typical workloads, \rev{it becomes less effective when 
an operator produces quantities (groups of tuples) too often, or the groups are small by nature. 
Hence, worst-case scenarios happen when operator states change frequently (for example, the state changes for each incoming tuple as in sliding windows) or when states (groups) contain very few tuples.}
\rev{In general, compression introduces a trade-off: while quantitative contributions are preserved, fine-grained interaction-level details (e.g., individual events, ordering within a window, or per-record lineage) are not retained. As a result, fine-grained provenance that requires such information (e.g., reconstructing the exact sequence of events leading to a specific output, or distinguishing between multiple interaction paths that contribute identically at the aggregate level) cannot be served by the TIN representation and compressed states alone.}
\rev{Finally, the accuracy of temporal attribution queries presented in Section \ref{sec:queries} relies on the time granularity of the aggregated quantities propagated through the dataflow modeled as a TIN. For temporal queries that ask for information at a finer granularity, the compressed information naturally can only give approximate estimates, in the same manner as histograms in selectivity estimation.}

%% file: solutions.tex
\vspace{-5pt}
\section{Discrete vs. Liquid Data}\label{sec:data}
Data classes 
\rev{may affect the options for temporal attribution and the dependencies between states.}
\textit{Discrete data} (\rev{data items} with unique identities) allows 
straightforward path-based provenance, while \textit{liquid data} 
(\rev{indistinguishable} quantities that merge and split) \rev{add challenges to temporal attribution mechanisms}.

\looseness=-1
\stitle{Discrete, Identifiable Data.} It consists of individual \rev{items (e.g., tuples)} that maintain a unique 
identity throughout their lifecycle.
\rev{While these items are 
transferred and transformed between workflow nodes, it is  possible
to track tuple-level dependencies, as streaming provenance systems do \cite{DBLP:conf/debs/GlavicEFT13, palyvos2018genealog, DBLP:journals/pvldb/Palyvos-Giannas20}}.
\rev{Hence, fine-grained provenance mechanisms can be used for temporal attribution, however, at a higher cost compared to modeling the dataflow as a TIN that captures transferred quantities instead of individual items.} 

\stitle{Liquid Data.} \rev{It refers to quantities that are not identifiable and cannot be tracked after transformations such as splitting, merging, etc.} \rev{In streaming dataflows, liquid data arises when numeric values are extracted from tuples and merged into aggregate totals. After being redistributed across operators, these values become \textit{un-named} quantities within a single sum; the resulting output obscures the origin of each unit, mirroring the ambiguity of merged funds in financial networks.} 
\rev{In liquid data, 
because 
the origin of an individual output item 
becomes ambiguous
and quantitative lineage becomes possible only for quantities (groups) of output items.}
\rev{As an analogy}, an amount of money, originating 
from one account, can be split across several transactions, merged with other 
funds, and eventually appear in multiple destinations. 
\rev{In scenarios where unidentifiable tuples are propagated and transformed fine-grained provenance is not even possible (e.g., when explicit links between items at distributed dataflows cannot be established, due to security or privacy constraints \cite{zhang2017privacy,politou2022privacy,pan2023data,zhang2023differentially}), so lineage can only be tracked quantitatively, by means of 
temporal attribution.
}

\rev{When modeling a dataflow as a TIN, we may think groups of transferred data as
liquid because 
we track aggregated  counts rather than individual items.} 
While each item may have a unique ID in the actual system, the TIN \rev{models the dataflow} 
at the flow aggregation level, \rev{as explained in the previous section}. 
\rev{Attribution meta-data do not differ between identifiable and liquid data, as long as the operators can identify how many tuples from a given input at time $t$ contributed to how many tuples to a specific output at time $t+1$. If liquid data consist of anonymized items, this is possible for all operators. In case where this is not possible (i.e., inputs are merged into indivisible quantities before the operator can process them), contribution propagation assumptions should be applied as in \cite{DBLP:conf/icde/KosyfakiM22}.}

%% file: temporalindex.tex
\vspace{-5pt}
\section{Temporal Provenance Indexing}\label{sec:tpi}
\rev{After modeling the dataflow as a TIN},
\rev{to} support diverse \rev{temporal attribution tasks}, we 
envisage indexing methods 
that capture both the state 
evolution of \rev{dataflow} vertices and \rev{attribution information on them} over time.
The key 
insight behind our approach is that each vertex in a TIN goes through a 
sequence of states, where each state corresponds to a time interval during 
which the vertex's 
\rev{quantities and their dependencies}
remain unchanged.
By temporally indexing these states, we can answer temporal attribution questions 
without \rev{replaying the 
dataflow}.

\noindent\textbf{State-based Representation.} Each vertex maintains a 
sequence of states characterized by time intervals, \rev{quantities}, 
and \rev{attribution} information. New states are created \rev{at each time window}
or when internal operations occur (e.g., windows in stateful operators fire),
naturally compressing temporal evolution by grouping periods with 
identical buffer states.

\noindent\textbf{Compression.} 
The state-based representation 
provides substantial compression compared to 
storing fine-grained contribution graphs at the tuple level. 
The key insight is that we create new 
states only when quantities \rev{at operators (and their dependencies)}
change.


\noindent\textbf{\rev{Attribution Computation}.} 
Each state maintains \rev{attribution} information that \rev{captures how much each of the states at previous operators contributes to the} 
current \rev{quantity at the state}.
\rev{Attribution can be computed by BFS along these state-to-state links}.
\rev{We may also explicitly propagate attribution information at states} as tuples of the form 
(origin, time, quantity), indicating that a specific quantity originated 
from a particular source vertex at a given time (or time interval). 
This 
allows us to \rev{know the origins/sources} of quantities back through the network \rev{and their attribution} in a compact representation. 
When a vertex transfers quantities outward, we \rev{propagate and} update 
the \rev{corresponding attribution data}.

\noindent\textbf{Index Structure.} 
The temporal \rev{attribution}  index \rev{(TAI)} organizes states chronologically for each 
vertex, enabling efficient \rev{time-based} retrieval of \rev{their buffered quantities} and \rev{attribution data}.
For a given vertex $v$ and time $t$, we can locate the 
relevant state using a B-tree over the temporal sequence of states, as they 
are naturally ordered by time. For example, to query $W_1$'s 
\rev{attribution}  at $t=3.5$, we perform search over its state sequence and 
retrieve state $s_2$, which covers the interval $[3,4)$. The index should support 
different query types (e.g., ``\rev{How much did each source contribute} at time $t$?'' or 
``How did \rev{attribution} evolve between $t_1$ and $t_2$?''). 
\rev{Temporal attribution queries use TAI to find the state(s) of the target  vertex that satisfy the temporal query predicate and then 
(temporally) join them with the 
states of adjacent vertices along the dataflow.}
Index maintenance 
occurs incrementally as new \rev{TIN interactions take place}. When an 
interaction modifies a vertex's buffer, we close the current state and 
create a new one, updating both the buffer quantity and the \rev{attribution} 
information based on the interaction.

\subsection{What Can We Ask?}\label{sec:queries}

We formalize five provenance query types that leverage temporal states in 
TINs and  illustrate them using the Flink pipeline from Figure~\ref{fig:overview}. \rev{This query set is not meant to be complete; our goal is to introduce some tools that can facilitate attribution-based analysis.}

\noindent\textbf{Q1: Backward \rev{Attribution} (Where-From).} 
Given \rev{sink} node \rev{$s_i$} \rev{and} time $t$, return all $\langle$source, time, 
quantity$\rangle$ tuples showing which origins \rev{(and how much)} contributed to $d$'s 
state \rev{at time $t$}.


\emph{Example:} ``At $t=4$, $W_1$ has 2000 events. What is the \rev{attribution of sources to them}?'' 
Trace backward: $M_1 \to 450$, $M_2 \to 775$, $M_3 \to 775$ (at $t=3$). 
Recursively: $S_1 \to M_1$ (900), $S_2 \to M_2$ (600), $S_2 \to M_3$ (775) 
at $t=2$, and ultimately $K_1, K_2, K_3$ at $t=1$. 

\noindent\textbf{Q2: Forward  \rev{Attribution} (Where-To).} 
Return all downstream destinations that \rev{
were affected by the input $s$ at time $t$, and how much was each of them affected.}

\emph{Example:} ``At $t=1$, $K_1$ ingests 1500 events. Where do they go?'' 
\rev{All events affect the Sink via path}: $K_1 \to S_1 \to M_1 \to W_1 \to$ Sink.
Forward provenance supports 
impact analysis.

\noindent\textbf{Q3: Temporal Lineage (When-Contributed).} 
Given vertex $v$ and time window $[t_1, t_2]$, return all sources whose 
contributions arrived \rev{at $v$} during that period.

\emph{Example:} ``Which sources contributed to $W_1$ between $t=2$ and 
$t=3$?'' At $t=3$: $M_1$ (450), $M_2$ (775), $M_3$ (775).


\noindent\textbf{Q4: Flow Lineage (How-Much-Through).} 
Given source $s$, \rev{sink $s_i$}, and intermediary $v$, compute the 
quantity that flowed from $s$ to $d$ via $v$ \rev{during time $[t_1,t_2]$}.

\emph{Example:} ``How much $K_1$ data reached $W_1$ via $M_1$ \rev{in $[3,4]$}?'' Path: 
$K_1 \to S_1$ (1500), $S_1 \to M_1$ (900), $M_1 \to W_1$ (450). Answer: 450.

\noindent\textbf{Q5: Versioning \rev{Attribution} (How-Changed).} 
Given vertex $v$ and times $t_1,t_2$, where $t_1 < t_2$, \rev{compute the change in attribution data of $v$ from $t_1$ to $t_2$}.

\emph{Example:} ``How did $W_1$'s \rev{attribution} change from $t=3$ to $t=4$?'' 
At $t=3$: \rev{$q$:} $2000$, \rev{attribution:}$ \{M_1:450, M_2:775, M_3:775\}$. At $t=4$: \rev{$q$:} $0$, 
\rev{attribution:}$ \emptyset$. Delta: all sources depleted (window fired).

All five queries support tracing through recursive temporal 
state lookups. \rev{Unlike applying expensive fine-grained provenance mechanisms, modeling the dataflow as a TIN, keeping attribution data with each state, and temporal indexing of states can facilitate fast retrieval of attribution information.}

%% file: relatedwork.tex
\section{Related Work}\label{sec:relwork}
\rev{Temporal attribution computation is closely related to data provenance.} 
Data provenance is well-studied in the research community
\cite{johns2025tracking,langhi2025evaluating, DBLP:conf/icde/KosyfakiM22,
  chapman2008efficient, chapman2020capturing,
  mohammed2025lineage,wang2017qfix, alabi2025privacy,
  gundecha2013seeking, taxidou2018information, barbier2013provenance,
  baeth2019detecting, taxidou2015modeling, davidson2013propagation,
  wu2019provcite, zhang2017privacy,
  reha2023anomaly, perez2018systematic, karvounarakis2010querying, DBLP:conf/debs/GlavicEFT13}. We briefly survey related work and position our contributions.

\stitle{Database Provenance.} Buneman et
al. \cite{DBLP:conf/icdt/BunemanKT01, DBLP:conf/pods/BunemanKT02}
introduced \linebreak where/why
provenance;
Green et al. \cite{green2007provenance,  green2017semiring} developed the semiring framework.
Surveys \cite{senellart2019provenance, tan2007provenance,
  glavic2021data} provide
comprehensive overviews. ProvSQL \cite{senellart2018provsql, widiaatmaja2025demonstration}
implements
provenance tracking in PostgreSQL. While these approaches excel at
static query provenance,
they do not 
address continuous data flows or provide temporal indexing mechanisms. 

\stitle{Streaming Provenance.} 
\rev{Ariadne \cite{DBLP:conf/debs/GlavicEFT13} introduces operator instrumentation for fine-grained provenance tracking in data stream processing systems,
extending 
the Borealis stream processing engine \cite{DBLP:conf/cidr/AbadiABCCHLMRRTXZ05}.
An independent provenance-focused work, also named Ariadne \cite{papavasileiou2019ariadne}, 
presents a system for declaratively customizing provenane capture and querying in large-scale graph analytics on vertex-centric graph processing platforms. 
Genealog \cite{palyvos2018genealog} provides fine-grained backward provenance for data streaming by annotating every output tuple with a constant-size provenance token that encodes which source tuples contributed to it. 
Ananke \cite{DBLP:journals/pvldb/Palyvos-Giannas20} extends Genealog 
to deliver forward provenance; a continuously maintained graph that records, for each source tuple, which downstream outputs it has contributed to so far and whether the source tuple can still influence future outputs. \cite{DBLP:journals/pvldb/Palyvos-Giannas22} is the first framework for why-not provenance in data streaming, explaining why a particular output was produced or why an expected output was absent by identifying the minimum subset of source data responsible. \cite{DBLP:journals/pacmmod/YamadaKSAM25} suggests a lazy provenance mechanism based on 
checkpointing operator state;  when  provenance is requested, the
dataflow is replayed from the nearest checkpoint, avoiding the throughput degradation of eager approaches.
All of these systems track {\em per-tuple provenance}: given an output tuple, they identify the set of source tuples that produced it. On the other hand, our vision is to add temporal attribution functionality to streaming systems, where the objective is to  measures the contribution of each source to a given output at a coarse level (e.g., all tuples generated within a time window). 
For this, the dataflow is modeled as a TIN where quantities (e.g., groups of tuples) flow together with lightweight attribution meta-data. 
Additionally, none of these systems
defines our temporal query types (backward/forward how-much, versioning, flow lineage, path-based provenance) and maintains a respective temporal index.}

\stitle{\rev{Provenance data compression.}}
\rev{Compression has been used within provenance frameworks. 
For example, Genealog \cite{palyvos2018genealog} replaces explicit lineage graphs with compact per-tuple annotations. 
Other approaches \cite{ainy2015approximated,lee2017integrating} generate concise representations by trading accuracy for compactness (e.g., via pattern-based or top-k summaries). 
In distributed settings, provenance compression techniques \cite{chen2017distributed} reduce storage and communication costs by compacting provenance graphs across nodes. Differently from all these approaches, our proposal incorporates compression directly into the data model via temporal state aggregation, representing flows and attribution at a coarser level while still supporting temporal attribution queries without reconstructing fine-grained lineage.}


\stitle{Temporal Provenance.} TAP/DTaP \cite{zhou2012distributed,
  zhou2011tap}
use distributed Datalog to
capture temporal provenance for distributed protocol debugging. Zeno
\cite{wu2019zeno} diagnoses performance problems using temporal
provenance.
We differ fundamentally by: 
(1) applying TINs to structurally model temporal provenance with 
explicit quantity tracking in data management contexts, 
(2) distinguishing discrete vs. liquid data classes that require 
different provenance semantics, (3) achieving compression via 
state-based indexing that groups consecutive time periods with 
identical buffer states, and (4) providing five temporal query types 
(backward, forward, temporal lineage, flow lineage, versioning) that 
leverage TINs' temporal structure.

\stitle{Graph Provenance.} Several efforts \cite{acar2010graph,
  kumar2017information, moreau2011open} model provenance
using
graph structures but typically represent static relationships. Prior
work on TINs \cite{DBLP:conf/edbt/KosyfakiMPT19,DBLP:conf/icde/KosyfakiMPT21,DBLP:conf/icde/KosyfakiM22}
examines flow computation and provenance tracking for liquid data in
TINs; \rev{we extend this with the definition of temporal attribution, state-based compression, provenance tracking
in stream networks, indexing for provenance, and support for a wide
range of temporal attribution queries}.

Our approach differs from all prior work by combining three elements: 
\circled{1} explicit \rev{attribution} tracking in \rev{dataflows} and temporal graphs; \circled{2} state-based compression achieving huge storage 
reductions for windowed workloads; and \circled{3} a query model 
supporting five temporal provenance query types (backward, forward, temporal 
lineage, flow lineage, versioning). To our knowledge, no existing
system provides all these capabilities together.

%% file: next.tex
\vspace{-5pt}
\section{The Path Forward} \label{sec:next}
Our vision of TIN-based temporal \rev{attribution}
opens several 
research 
directions. 
We organize our agenda around three core challenges.

\stitle{\textit{\rev{Attribution} Query Optimization challenge}:}
\rev{Temporal attribution queries (Section \ref{sec:queries}) can be modeled as database queries over temporal tables that store attribution meta-data for each state. 
To obtain the attribution graph for a particular state, we should  perform temporal joins with the states of previous nodes. Hence, a query optimization challenge arises.}

\textbf{Research Questions:} \circled{1} What \rev{attribution}-specific statistics 
(state transition frequency, \rev{state dependencies}, temporal correlation) 
\rev{can be used for} query planning? \circled{2} When can optimizers rewrite recursive 
provenance queries to skip deep traversal? For instance, a backward 
provenance query with depth limit $k$ only requires the most recent 
$k$ state transitions, avoiding historical states. \circled{3} How should 
systems handle skewed provenance distributions where some vertices 
have hundreds of contributing sources?

\textbf{Approach:} Represent state-based indices as relational tables, 
enabling  engines like DuckDB
to leverage vectorized execution 
and compression. Design tiered storage (hot states in memory, warm on 
SSD, cold in object storage) balancing latency and cost.


\stitle{\textit{Adaptive Compression and Unified Models challenge}:}
State-based compression effectiveness varies 
dramatically across workloads; windowed streaming achieves million-fold 
compression while high-frequency trading sees diminished benefits. 

\textbf{Research Questions:} \circled{1} How does query performance vary \rev{with time granularity used in} compression? \circled{2} \rev{Can we predict optimal compression 
granularity for a given workflow and attribution requirements?}
\circled{3} When should systems transition 
between \rev{fine-grained provenance } and \rev{temporal attribution}?

\textbf{Approach:} Develop learned compression policies that observe 
workload characteristics (query temporal resolution, update rates, \rev{attribution analysis requirements}) and 
dynamically adjust state granularity per vertex. 
Design unified models 
supporting both \rev{fine-grained provenance} and \rev{attribution}, 
enabling automatic conversion at operator boundaries.


\stitle{\textit{Distributed \rev{Attribution} Indexing challenge}:}
\rev{In distributed workflows that process big data tasks}, centralized storage \rev{and indexing of temporal attribution data may not be feasible.} 

\textbf{Research Questions:} \circled{1} What partitioning strategies (by vertex 
ID, time ranges, or query patterns) minimize cross-\rev{machine} \rev{temporal attribution} queries? 
Time-based partitioning enables efficient temporal queries but splits 
high-degree vertices across partitions, while vertex-based partitioning 
co-locates \rev{attribution data for neighboring nodes}, but complicates time-range queries. Can hybrid 
(vertex, time\_bucket) partitioning balance both? \circled{2} What consistency 
models balance staleness versus overhead? \circled{3} How can vector clocks 
maintain causal ordering across replicas?

\textbf{Approach:} Extend states with causal metadata, design gossip 
protocols for asynchronous synchronization, and develop partition-aware 
query planners.



%% file: conclusions.tex
\section{Conclusions}\label{sec:conclusions}
Fine-grained provenance mechanisms for streaming systems face scalability challenges as contribution graphs grow superlinearly 
with data volume.
\rev{When it comes to analyzing historical data, the problem escalates, as we need to maintain fine-grained provenance meta-data over a long history.}
To alleviate this, we define the concept of \rev{temporal attribution for modeling the impact of inputs (sources) to the output in dataflows at a coarser, quantitative level compared to tuple-level provenance.}  
\rev{We suggest}
Temporal 
Interaction Networks (TINs) \rev{as a way} to
\rev{model data flow at a coarser, compressed level over time}.
\rev{At the same time, we suggest the propagation of temporal attribution meta-data together with the flow}
capturing structural 
and temporal aspects for richer analysis. Through examples in streaming \rev{workflows},
we classify data into identity-preserving and \rev{liquid} 
types, define five temporal query types, and propose a state-based index for 
compressed querying. Our vision opens research directions in \rev{query optimization for temporal attribution analytics}, adaptive compression, unified models for diverse data, and 
distributed indexing 
and advances temporal provenance
from a debugging tool to a 
broad data management primitive.